\documentclass[aps,pre,twocolumn,showpacs,superscriptaddress,groupedaddress]{revtex4}
\usepackage{graphicx}  
\usepackage{dcolumn}   
\usepackage{bm}        
\usepackage{amssymb}   
\usepackage[english]{babel}
\usepackage{color}
\newcommand{\ev}{\, \mathrm{eV}}
\newcommand{\tr}{\, \mathrm{Tr}}
\newcommand{\gcc}{\, \mathrm{\frac{g}{cm^3}}}

\makeatletter

\@ifundefined{textcolor}{}
{%
 \definecolor{BLACK}{gray}{0}
 \definecolor{WHITE}{gray}{1}
 \definecolor{RED}{rgb}{1,0,0}
 \definecolor{GREEN}{rgb}{0,1,0}
 \definecolor{BLUE}{rgb}{0,0,1}
 \definecolor{CYAN}{cmyk}{1,0,0,0}
 \definecolor{MAGENTA}{cmyk}{0,1,0,0}
 \definecolor{YELLOW}{cmyk}{0,0,1,0}
}

\makeatother

\usepackage{babel}
\begin{document}

\title{The Leading Correction to the Thomas-Fermi Model at Finite Temperature}

\date{\today}

\author{Eyal Segev }

\email{segev.eyal@gmail.com}

\author{Doron Gazit}

\email{doron.gazit@mail.huji.ac.il}

\affiliation{Racah Institute of Physics, Hebrew University, 91904 Jerusalem, Israel}

\begin{abstract}

The semi-classical approach leading to the Thomas-Fermi (TF) model provides a simple universal thermodynamic description of the electronic cloud surrounding the nucleus in an atom. This model is known to be exact at the limit of $Z\rightarrow\infty$, i.e., infinite nuclear charge, at finite density and temperature. Motivated by the zero-temperature case, we show in the current paper that the correction to TF due to quantum treatment of the strongly bound inner-most electrons, for which the semi-classical approximation breaks, scales as $Z^{-1/3}$, with respect to the TF solution. As such, it is more dominant than the quantum corrections to the kinetic energy, as well as exchange and correlation, which are known to be suppressed by $Z^{-2/3}$. We conjecture that this is the leading correction for this model. In addition, we present a different free energy functional for the TF model, and a successive functional that includes the strongly bound electrons correction. We use this corrected functional
to derive a self-consistent potential and the electron density in
the atom, and to calculate the corrected energy. At this stage, our model has a built-in validity limit, breaking as the L shell ionizes.
\end{abstract}
\maketitle
\section{Introduction}
The field of Hot or Warm and Dense Matter (HDM/WDM) is of central importance in astrophysics, where the interior of the Sun, as well as other main 
 sequence stars, is composed of similar atomic plasma. In recent years, extreme thermodynamic conditions are achieved terrestrially using large scale experimental facilities,  
 such as the Z-machine at Sandia National Laboratories, or the National Ignition Facility (NIF), where plasma can be produced at local thermodynamic equilibrium with 
 temperatures of the order of $300\ev$. 
Tackling  the many-body problem  of  atoms in these conditions is difficult, despite the simplicity of the underlying coulomb potential. In order to solve this problem, one resorts to different approximations of the many-body quantum problem. A different approach, whose foundations lie in Density 
Functional Theory (DFT), is to formulate the problem in terms of the mean electron density. This is particularly useful when studying thermodynamic properties of a gas of atoms, at 
finite temperature and density. 

The first example of a density functional formulation of the atom has in fact been achieved long before the development of DFT in the Thomas-Fermi (TF) model \cite{2,3}. 
It is based on a semi-classical treatment of the electrons in the atom. Soon after its derivation, the zero-temperature model has been corrected to include quantum gradient corrections to the kinetic 
energy and exchange effects \cite{4,5}. Scott, and following studies, have investigated the limits of the semi-classical approximation, by separating the strongly bound electrons from the semi-classical integration \cite{6}. The basic TF model, as well as the gradient and exchange corrections, were generalized to finite temperature and densities.\cite{18,19,20,21,30}. These generalizations were combined with the ion-sphere model. In this model, the atom is enclosed in a spherically
symmetric cell that contains all the electrons of the atoms 
to ensure neutrality. Such models define an electron-density
distribution that obeys self-consistency, assuming exact cancellation of the free-electron and other
ion densities beyond the sphere \cite{200,201,102}. The TF model 
is a very crude and basic approximation for the atom, but its foundations are basic principles of physics, and in fact it is exact at $Z 
\rightarrow \infty$ \cite{33,51,34,35}, with $Z$ the number of protons in the nucleus of the atom. Moreover, all physical properties predicted by TF model have a simple scaling property with $Z$. 
TF model thus provides a universal description of all materials, differing only by a scaling factor.

The development of Kohn-Sham DFT (KS-DFT) \cite{8,9,39} has highlighted the advantages of the TF model. KS-DFT ensures that the ground state properties of a quantum many body system are dictated by its 
density. However, KS-DFT does not hint towards the structure of the density functional that governs the system properties.
As a result, TF model is commonly used as the limit for phenomenological DFT models of heavy atoms \cite{50}, as TF depends merely on densities by construction . 
Moreover, Generalized Gradient Approximation (GGA) for the kinetic energy and an accurate exchange-correlation
term, have been derived as corrections of the finite temperature TF model using Orbital Free DFT (OFDFT) \cite{37,38}. 
A different approach to the corrections to TF model was taken by Schwinger and Englert \cite{12,10,13}. Studying a TF description of an isolated atom, i.e., a zero temperature model, they systematically ordered the corrections to the leading TF model by their $Z$ dependence. They 
demonstrated that Scott's correction, i.e., a quantum treatment of the strongly bound electrons, is the leading correction to TF model, suppressed by $Z^{-1/3}$. Other corrections,
such as quantum and exchange, are of lower order, suppressed by $Z^{-2/3}$. 

In the current paper we develop a generalization of Scott's correction to the finite temperature and density TF model. We show that it is more dominant than the known quantum and exchange-correlation corrections, and thus conjecture that it is the leading 
correction at these thermodynamic conditions. As such, our study is the initial step to establish a systematic order of density functionals, whose different predictions represent the uncertainty in predictions.
\section{FINITE TEMPERATURE THOMAS-FERMI}
\subsection{Orbital Free Density Functional Formalism}
We start by considering the assumptions leading to the TF model for a neutral atom of charge $Z$ at finite temperature $T$ and chemical potential $\mu$. The single electron Hamiltonian is 
 \begin{eqnarray} 
 H=\frac{p^{2}}{2m}+\left(-e\right)V(\vec{r}).
 \end{eqnarray} 
Neglecting exchange and correlations contributions, the single particle internal energy is
\begin{eqnarray} 
U_{1}=\tr\left( Hn_{FD}\left(r,p\right)\right),
\end{eqnarray}
where the electronic density is the Fermi-Dirac distribution. The total entropy can be calculated combinatorially to be
 \begin{eqnarray} 
 S=-k_{B} \tr\left(n_{FD} \ln\left(n_{FD}\right)+\left(1-n_{FD}\right)\ln\left(1-n_{FD}\right)\right),
 \end{eqnarray} 
where $k_B$ is Boltzmann constant.

Summing all the single particle energies leads to double 
counting of the electrostatic energy between the electrons, thus we subtract it once, and write the free-energy functional as \cite{11} 
\begin{eqnarray}
F&=&\tr\left(\left(H-\mu\right)n_{FD}\right)\nonumber-TS\\&-&\frac{1}{8\pi}\int d^{3}r \left(\vec{\nabla}\left(V(\vec{r})-\frac{eZ}{r}\right)\right)^{2}+\mu Z,
\end{eqnarray} 
when a term for the entropy was also added. The TF main approximation is to evaluate the free energy functional by taking a semi-classical trace over the states, i.e., $\tr \rightarrow \frac{2}{(2\pi\hbar)^{3}}\int d^{3}r d^{3}p.$
After $p$ integration, 
\begin{eqnarray}
F_{TF} &=&   \int d^{3}r \left(\frac{\sqrt{2}}{\pi^{2}\hbar{}^{3}}\frac{m^{\frac{3}{2}}}{\beta^{\frac{5}{2}}}\left(-\frac{2}{3}I_{\frac{3}{2}}(\beta(eV(\vec{r})+\mu))\right)\right. 
\nonumber \\&-&\left.\frac{1}{8\pi} \left(\vec{\nabla}\left(V(\vec{r})-\frac{eZ}{r}\right)\right)^{2}\right)+\mu Z \label{eq:2},
\end{eqnarray}
here $\beta=(K_BT)^{-1} $, and $I_{z}(\eta)$ is the Fermi-Dirac integral \footnote[1]{$I_{z}(\eta) = \int\frac{dxx^{z}}{exp(x-\eta)+1}.\quad z\geq-1$}.

 The advantage of writing the TF model in terms in this novel way is
that a variation of this functional leads to two self-consistent relations. A variation with respect to $\mu$ leads to the number constraint,
\begin{eqnarray}
Z=\int d^{3}r \, n(\vec {r}),
\end{eqnarray}
with
\begin{eqnarray} 
n(\vec{r})=\frac{\sqrt{2}}{\pi^{2}\hbar{}^{3}}\frac{m^{\frac{3}{2}}}{\beta^{\frac{3}{2}}}I_{\frac{1}{2}}\left(\beta\left(eV(\vec{r})+\mu\right)\right).
\end{eqnarray}
Variation with respect to the potential leads to 
Poisson equation
\begin{eqnarray}
\nabla^{2}\left(V(\vec{r})-\frac{eZ}{r}\right) & = & 4\pi en(\vec{r}).\label{eq:49}
\end{eqnarray}
These two equations of the TF model at finite temperature and density are solved as in \cite{18,19,30}, using the approximated ion-sphere model, by setting a finite radius to the atom $r_0$ and determining the following boundary conditions $V(r_{0})=0$ and $\quad\frac{dV(r)}{dr}\mid_{r=r_{0}}=0$. The cell radius $r_0$ is determined by the Wigner-Seitz cell, i.e.,
\begin{eqnarray}
r_{0}  =  \left(\frac{3}{4\pi}\frac{A}{\rho N_{A}}\right)^{\frac{1}{3}},
\end{eqnarray}
where $N_{A}$ is the Avogadro number $A$ is the atomic mass and $\rho$ is the plasma density. We also assume a point nucleus at the center of the atom with the boundary condition $rV(r)|_{r=0}=eZ$. 
\subsection{Scaling}
We turn our attention to a new way to extract the scaling of the TF energy in $Z$.
In this subsection we use Hartree atomic units. We define
\begin{eqnarray}
F_{TF} & = & F^{(1)}+F^{(2)}+\mu N,\label{eq:66}
\end{eqnarray}
where
\begin{eqnarray}
F^{(1)} & =- & \int d^{3}r\frac{2\sqrt{2}}{3\pi^{2}\beta^{\frac{5}{2}}}I_{\frac{3}{2}}\left(\beta\left(V(\vec{r})+\mu\right)\right),\label{eq:67}
\end{eqnarray}
and 
\begin{eqnarray}
F^{(2)} & = & -\frac{1}{8\pi}\int d^{3}r\left(\nabla\left(V(\vec{r})-\frac{Z}{r}\right)\right){}^{2}.\label{eq:22}
\end{eqnarray}

We rescale: $r\rightarrow\lambda^{-1}r$ $V(r)\rightarrow\lambda^{\alpha}V(\lambda r)$
,$\beta\rightarrow\lambda^{-\alpha}\beta$ $r_{0}\rightarrow r_{0}\lambda^{-1}$.
In order to comply with the boundary condition at $r=0$, which yields
$rV(r)|_{r=0}=eZ,$ we must have $Z=\lambda^{\alpha-1}Z$ . In addition,
we rescale $\mu$ in the following manner $\mu\rightarrow\lambda^{\alpha}\mu$
for later convenience. For the neutral atom case, $N=Z$, we have
\begin{eqnarray}
F^{\lambda}_{TF} & = & \left(\lambda^{\frac{5}{2}\alpha-3}F^{(1)}+\lambda^{2\alpha-1}F^{(2)}\right)+\lambda^{2\alpha-1}\mu Z.\label{eq:68}
\end{eqnarray}
We first note that
\begin{eqnarray}
\frac{dF^{\lambda}}{d\lambda}\mid_{\lambda=1} & =\nonumber \\
& = & \frac{\partial F^{\lambda}}{\partial Z}\frac{d}{d\lambda}\left(\lambda^{\alpha-1}Z\right)\mid_{\lambda=1}\nonumber+\nonumber\frac{\partial F^{\lambda}}{\partial \beta}\frac{d}{d\lambda}\left(\lambda^{-\alpha}\beta\right)\mid_{\lambda=1}\nonumber\\
& &+ \nonumber\frac{\partial F^{\lambda}}{\partial r_0}\frac{d}{d\lambda}\left(\lambda^{-1}r_0\right)\mid_{\lambda=1}\nonumber \\
& = & (\alpha-1)Z\frac{\partial F}{\partial Z}-\alpha\beta\frac{\partial F}{\partial\beta}-r_{0}\frac{\partial F}{\partial r_{0}}\label{eq:302}.
\end{eqnarray}

This is a partial differential equation we need to solve in order
to get $F_{TF}=F_{TF}(Z,\beta,r_{0})$. If one chooses $\alpha=4$, the RHS of Eq. (\ref{eq:68}) is homogeneous in $\lambda^{7}$. Thus, Eq. (\ref{eq:302}) is simplified
\begin{eqnarray}
7F & = & 3Z\frac{\partial F}{\partial Z}-4\beta\frac{\partial F}{\partial\beta}-r_{0}\frac{\partial F}{\partial r_{0}}.\label{eq:69}
\end{eqnarray}
Using the characteristics method we get
\begin{equation}
F=Z^{\frac{7}{3}}f(\sigma,\tau),\label{eq:20}
\end{equation}
where $f$ is the universal function of the TF model and,
\begin{equation}
\sigma=\frac{1} {r_{0}Z^{\frac{1}{3}}},\tau=\frac {1}{Z^{\frac{4}{3}}\beta}.\label{eq:205}
\end{equation}
When $\sigma \rightarrow 0 $ and $\tau \rightarrow 0 $, we recover that
\begin{equation}
f(\sigma,\tau) \rightarrow -0.768745,
\end{equation}
which is the known zero temperature result.

In order to get this result we must have
\begin{eqnarray}
V_{TF}= \tilde{V}(\tilde{r},\sigma,\tau)Z^{\frac {4}{3}}, \label{2}
\end{eqnarray}
when $\tilde{V}$ is the scaled potential $r=\tilde{r} Z^{\frac {1}{3}}$. We also must have
\begin{equation}
\mu=c\left(\sigma,\tau\right)Z^{\frac{4}{3}} \label{1},
\end{equation}
where $c$ is the universal coefficient of the chemical potential.
 In order to expand V close to the nucleus, we define
\begin{eqnarray}
\Phi(\frac{r}{r_0})\equiv\frac{\beta}{r_0}r\left(eV(r)+\mu\right). \label{3}
\end{eqnarray}
We expand $\Phi$ up to the first order as a series around zero. In order to get a better expansion we need to turn to Baker's expansion due to the fact that $\Phi''(\frac {r}{r_0})$ is not continuous at $r=0$ \cite{28}.
In the first order we have
\begin{eqnarray}
V(r)\cong V_B(r)\equiv-\frac{eZ}{r}+B(\sigma,\tau)Z^{\frac {4}{3}}. \label{4}
\end{eqnarray}
When $\sigma \rightarrow 0 $ and $\tau \rightarrow 0 $, we recover that 
\begin{equation}
B(\sigma,\tau) \rightarrow 1.793.
\end{equation}
\section{The Strongly Bound Electrons at Finite Temperature }
\subsection{Analytical Derivation}
One of the shortcomings of the TF model are unphysical properties near the nucleus. This is a result of the fact that the semi-classical approximation breaks near the nucleus, as the electrons wavelength becomes comparable to the distance to the nucleus. This effectively 
imposes an ultraviolet cutoff $\mu_{s}$ on the semi-classical trace. However, such a cutoff neglects the energy of strongly bound electrons. 

We incorporate the strongly bound electrons perturbatively into the TF
model by subtracting the semi-classical summation above the
ultraviolet cutoff, and adding the quantum mechanical trace of the
strongly bound electrons. We setup a consistent perturbative
calculation, thus calculating the trace for the strongly bound
electrons using the leading order potential, i.e., the potential
resulting from the TF model. As the inner electrons are strongest bound, we use Eq. (\ref{4}), i.e., a coulomb potential, leading to hydrogen atom wave functions and energies, up to the constant $B Z^{\frac{4}{3}}$. This approximation is valid for a shell $j$ with $n_j$ electrons, if
\begin{eqnarray}
-\frac{Z^{2}me^{4}}{2\hbar^{2}n_{j}^{2}}+BZ^{\frac{4}{3}}\gg \left< eV(r)-eV_B(r)\right>_{n_j}.
\end{eqnarray} The deviation is at the percentage level for K-shell electrons, and about $10\%$ for L-shell electrons. 
Finally,
\begin{eqnarray}
F_{TFS} & = & F_{TF}-F_{\mu_{s}}+\left\langle \frac{p^{2}}{2m}-\frac{e^{2}Z}{r}\right\rangle_{n_s} \nonumber \\
 &  & +\left\langle \frac{e^{2}Z}{r}-eV(r)-\mu\right\rangle_{n_s} ,\label{eq:79}
\end{eqnarray}
we define $n_s$ to be the quantum number of the highest
energy level smaller than $\mu_s$.
In addition, we introduced $F_{\mu_{s}}$, the semi-classical subtraction of the
energy of particles whose energy is smaller than $\mu_{s}$. We use $n_{\tilde{FD}}$
the Fermi-Dirac distribution with chemical potential $\mu_{s}$
in order to select the electrons with energy less than $\mu_{s}$. We assumed that these electrons, which are very close to the nucleus, and strongly bound, are not influenced by the heat reservoir. Thus, we can treat them as if they were at zero temperature. We checked this numerically and saw no change in the result within the regime of validity of the presented model.

$\mu_s$ is chosen in the gap between the K-shell and L-shell, aiming
to use a quantum mechanical treatment for the K-shell, i.e.,
$\mu_{1}<\mu_{s}<\mu_{2}$ (therefore $n_s=1$), where
\begin{eqnarray}
\mu_{j}&=&-\frac{Z^{2}me^{4}}{2\hbar^{2}n_{j}^{2}}+\left\langle \frac{e^{2}Z}{r}-eV(r)\right\rangle _{n_{j}}.\label{eq:211}
\end{eqnarray}
As a result, 
\begin{eqnarray}
F_{\mu_{s}} & = & \frac{1}{2}\sum_{j=1}^{2}F_{\mu_{j}}.\label{eq:1239}
\end{eqnarray}
Complying with the assumption that the strongly bound electrons fill complete shells, without fluctuations, dictates
$\mu_1<\mu_2\ll\mu$.

Taking the trace in Eq. (\ref{eq:79}) we deduce the energy
functional for TF model at finite temperature with the
strongly bound electron correction, Thomas-Fermi-Scott(TFS),
\begin{widetext}
\begin{eqnarray}
F_{TFS} & = & \int d^{3}r\left(\frac{\sqrt{2}}{\pi^{2}\hbar{}^{3}}\frac{m^{\frac{3}{2}}}{\beta^{\frac{5}{2}}}\left(-\frac{2}{3}I_{\frac{3}{2}}(\beta(eV+\mu))\right)-\frac{1}{8\pi}(\nabla(V(r)-\frac{eZ}{r}))^{2}\right)+\mu Z\nonumber \\
 &  & -\frac{1}{2}\Sigma_{j=1}^{2}\int d^{3}r\left(\frac{m^{\frac{3}{2}}\sqrt{2}}{\pi^{2}\hbar^{3}\beta^{\frac{5}{2}}}\left(-\frac{2}{3}I_{\frac{3}{2}}(\beta(eV+\mu_{j}))+\beta(\mu_{j}-\mu)I_{\frac{1}{2}}(\beta(eV+\mu_{j}))\right)\right)\nonumber \\
 &  & +\int d^{3}r\left(\frac{e^{2}Z}{r}-eV(r)-\mu\right)\rho_{s}(r)-\frac{Z^{2}me^{4}}{\hbar^{2}},\label{eq:140}
\end{eqnarray}
\end{widetext}
with
$\rho_{s}(r)\equiv 2\left|\psi_{1}(r)\right|^{2}$,
and $\psi_{1}(r)$ are Hydrogen-like ground state wave functions.

Similarly to the leading order, a variation of this functional with respect to $\mu$ and $V$ leads to the number constraint and to the Poisson equation. However, the electron density is now 
divided to the density of strongly bound electrons and the rest of the electrons,
\begin{eqnarray}
n(r) & = & \tilde{n}(r)+n_{strong}(r),\label{eq:147}
\end{eqnarray}
with
\begin{eqnarray}
n_{strong}(r) & = & \rho_{s}(r)+\frac{1}{2}\Sigma_{j=1}^{2}Q_{j}\left|\psi_{n_{j}}(r)\right|^{2},\label{eq:148}
\end{eqnarray}
when we define
\begin{eqnarray} 
Q_{j}  &=& \frac{dF_{\mu_i}}{d\mu_i} \\
&=&\int d^{3}r\frac{m^{\frac{3}{2}}\sqrt{2}}{2\pi^{2}\hbar^{3}\beta^{\frac{3}{2}}}\left(\beta(\mu-\mu_{j})I_{-\frac{1}{2}}(\beta(eV+\mu_{j}))\right) \nonumber,
\end{eqnarray}
and
\begin{eqnarray}
\tilde{n}(r) & = & \frac{\sqrt{2}}{\pi^{2}\hbar{}^{3}}\frac{m^{\frac{3}{2}}}{\beta^{\frac{3}{2}}}\left(I_{\frac{1}{2}}(\beta(eV+\mu))\right. \\ \nonumber &-&\left.
\frac{1}{2}\Sigma_{j=1}^{2}\left(\left(I_{\frac{1}{2}}(\beta(eV+\mu_{j}))\right)\right.\right.  \\ \nonumber &+&\left.\left.\left(\frac{1}{2}\beta(\mu_{j}-\mu)I_{-\frac{1}{2}}(\beta(eV+\mu_{j}))\right)\right)\right).\label{eq:149}
\end{eqnarray}
This correction, due to the strongly bound electrons, clearly does not change the chemical potential, since the correction does not change the dependence of the functional upon the number of electrons.

A lengthy analytical derivation in Appendix A, using the characteristics method, leads to the scaling properties of the model (in Hartree atomic units),
\begin{equation}
F_{TFS}=f_{TF}\left(\sigma,\tau\right)Z^{\frac{7}{3}}+\frac{1}{2}Z^{2}+O(Z^{\frac{4}{3}}),\label{eq:122}
\end{equation}
predicting that the prefactor to Scott's $Z^2$ term is density and temperature independent.
\subsection{Numerical Results}
\begin{figure}[h]
\includegraphics[ width=0.6\textwidth]{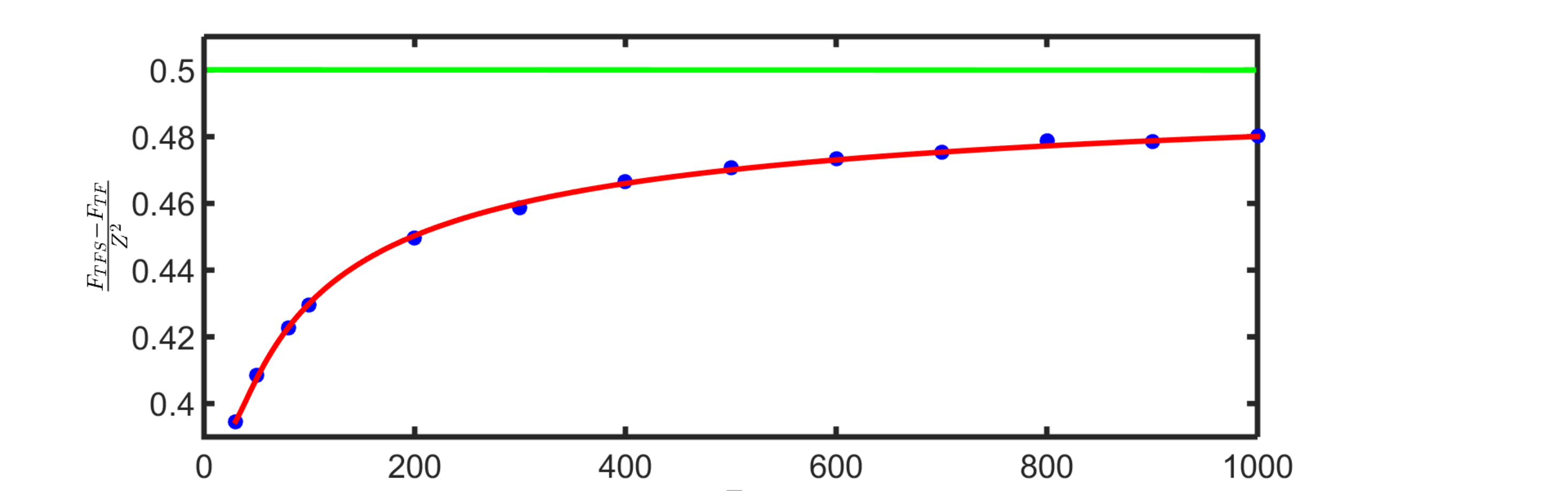}
\caption{The  $Z$ dependence of the Scott coefficient. The blue dots were numerically calculated, and the red curve  is a fit to the analytical dependence of this coefficient
  as in Eq. (\ref{eq:122}), showing asymptotic approach to Scott's value of
  $\frac{1}{2}Z^2$ (green curve).
 \label{fig:10}} 
 \end{figure}

\begin{figure}[h]
	\includegraphics[width=0.5\textwidth]{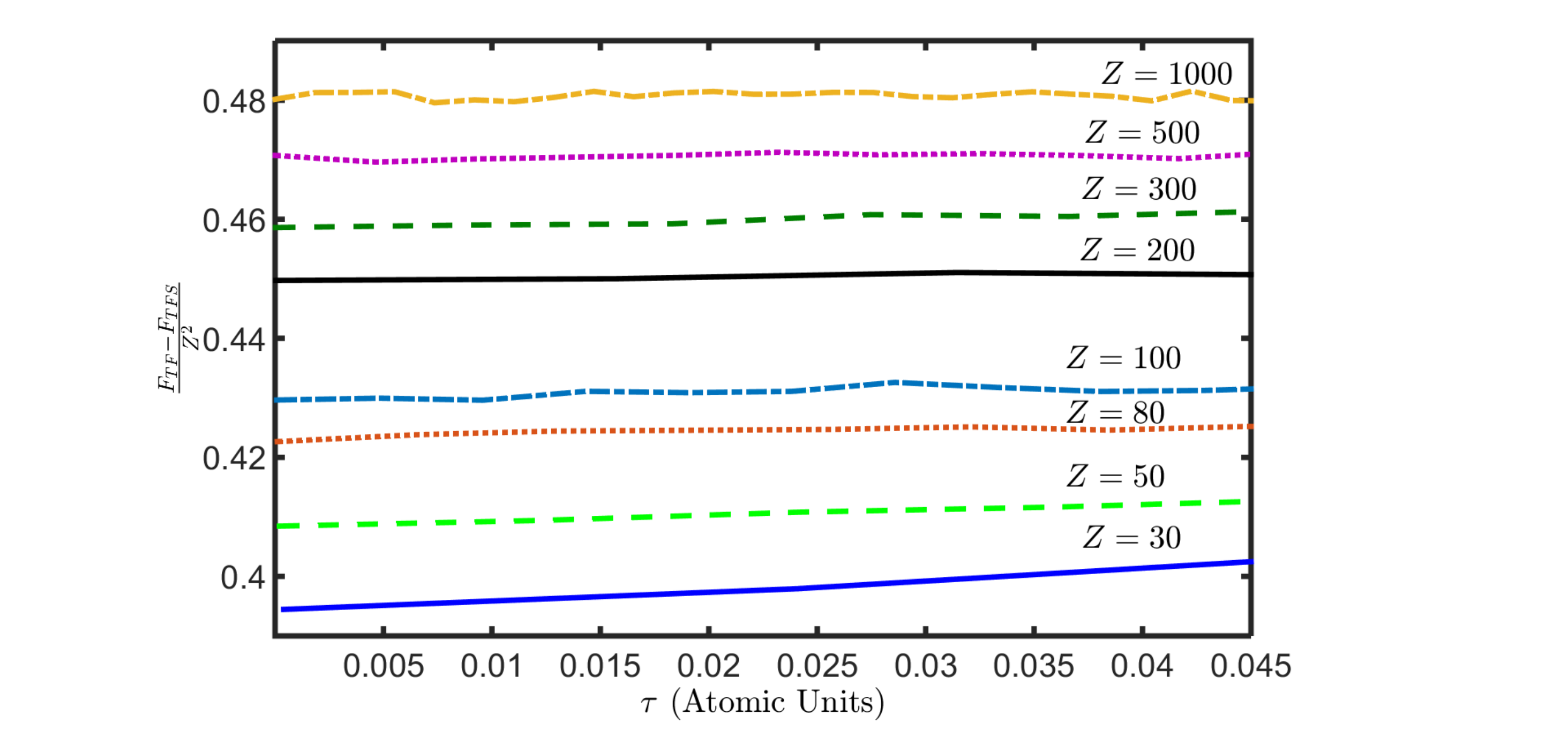}
	\caption{Dependence of the the Scott coefficient on $\tau$. Accomplished for
		$Z=30,\,50,\,80,\,100,\,200,\,300,\,500,\,1000$ and
		$\rho=1\gcc$. Essentially no dependence is found. Similar
		independence is found as a function of $\sigma$.
		\label{fig:9}} 
\end{figure}

Our model is verified numerically in Fig.~\ref{fig:10} and Fig. ~\ref{fig:9}, where we demonstrate the scaling as derived from the free energy, and invariant of the scaled temperature $\tau$. Similar constant behavior is verified for the (scaled) density $\sigma$. This is due to the neglect of temperature and density effects on the K shell electrons. The scaling clearly shows that the generalized Scott's correction is suppressed by $Z^{-1/3}$, with respect to the bare TF result. Moreover, as $Z \longrightarrow \infty$, the coefficient goes to $\frac{1}{2}$ monotonically, validating our analytic derivation.

\begin{figure}[h]
\includegraphics[width=0.5 \textwidth]{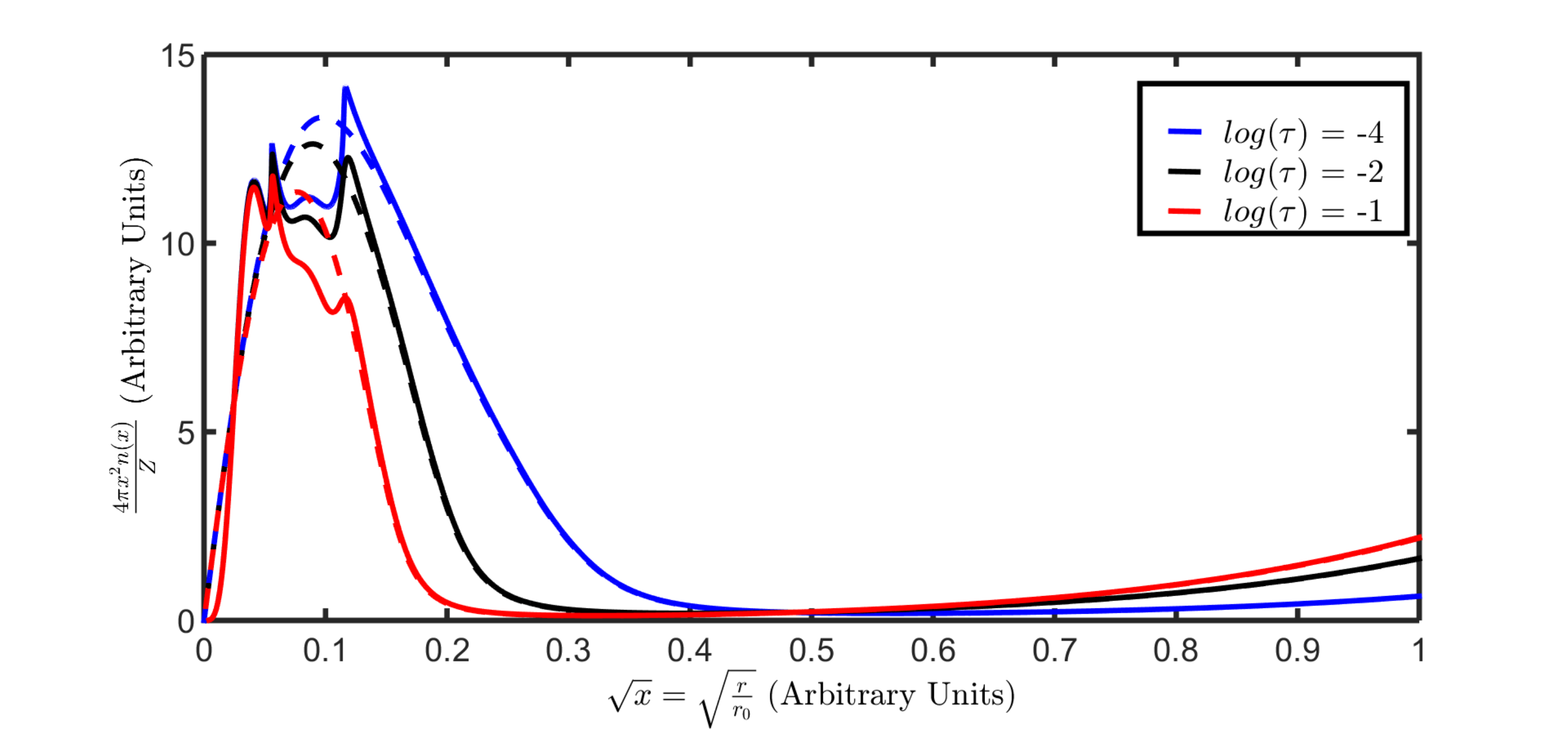}
\caption{Dependence of the scaled radial electron density $\frac{4\pi x^{2}n(x)}{Z}$ on $x=\sqrt{\frac{r}{r_0}}$
according to the TF (dashed lines) and the TFS (solid lines)  models. Shown  are explanatory calculations for  $Z=80$, $\rho=1\frac{g}{cm^{3}}$ at $T=1 \ev,\:500\ev,\:1000\ev$ or $\tau=1.06\times10^{-4},\:1.07\times10^{-2},\:1.166\times10^{-1}$ in Atomic Units. \label{fig:4}}
\end{figure}

\begin{figure}[h]
\includegraphics[width=0.5\textwidth]{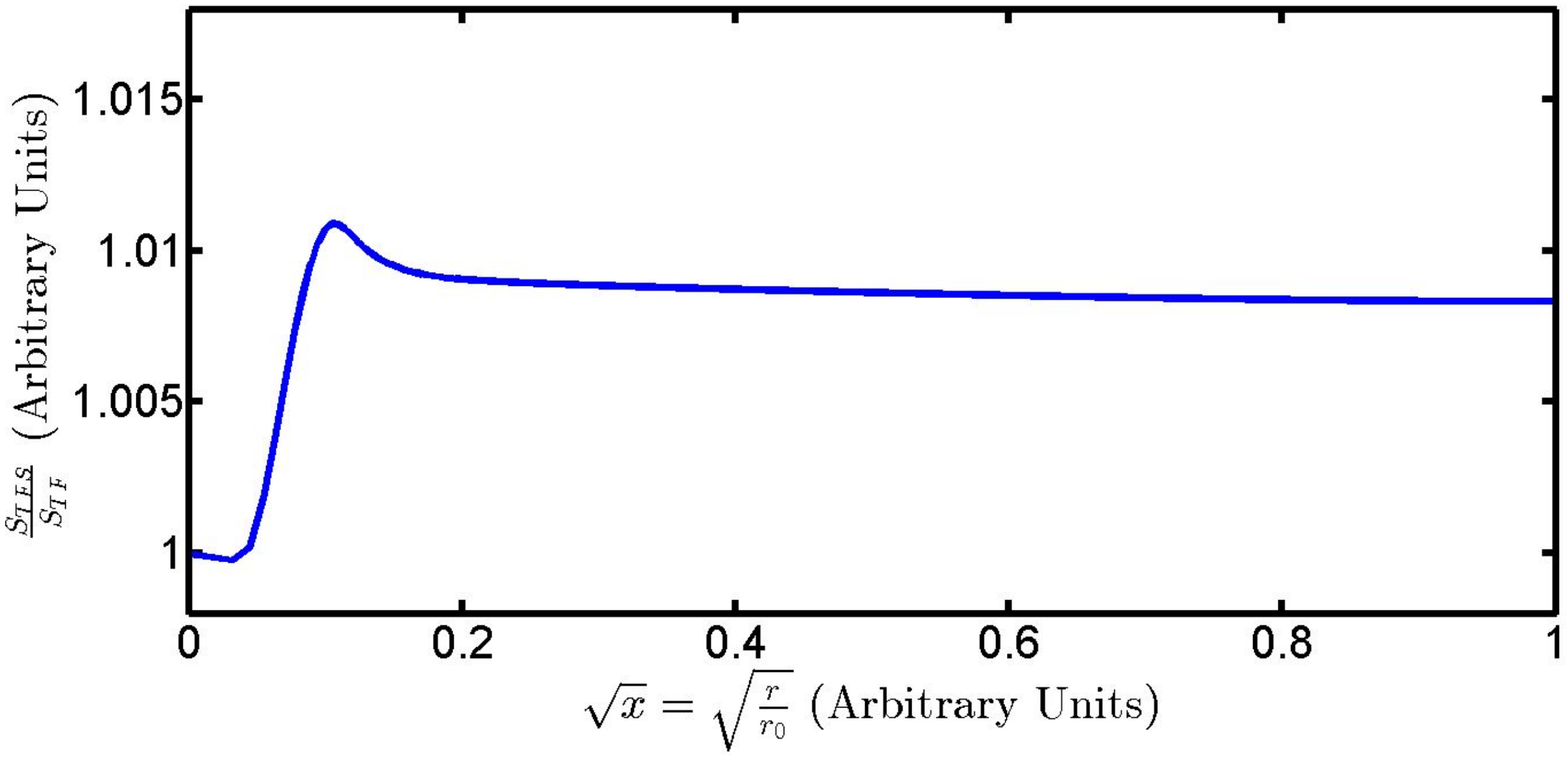}
\caption{The difference between the screening functions of the TF and TFS models.
Shown are calculations for Mercury $Z=80$, $\rho=1\frac{g}{cm^{3}}$ and
at $T=1000\ev.$\label{fig:5}}
\end{figure}

The TFS correction affects the atomic potential, and as a result the electronic densities. In Fig.~\ref{fig:4} and Fig.~\ref{fig:5} we demonstrate this for Mercury (Z=$80$) at $\rho=1\gcc$. Changes in screening factor, Fig.~\ref{fig:5}, reach up to 1\% near the nucleus, and diminish further from the nucleus, as expected. This reproduces the result for temperature zero in  Ref.~\cite{13}. The electronic densities show the expected near nucleus shell structure, and are temperature dependent. In Table I we show some numerical results of the parameters of the model. The low-temperature limit in Fig.~\ref{fig:4} and Table I reproduces the results of Ref.~\cite{10}.
\begin{table}
       \caption{Numerical values of the TFS model. Calculations for $Z=80$, and $\rho=1\frac{gr}{cm^3}$.}                       
    \begin{tabular}{ | l | l || l| l| l| l|}
    \hline                 
    T & $\tau$ & $\mu_1$ & $\mu_2$&$Q_1$ &$Q_2$ \\ \hline
    1eV & 0.0001 & -2724 & -443.8 & 0.867 & 5.126 \\ \hline
    10eV & 0.001& -2727 & -444.9& 0.866 & 5.125     \\ \hline
    100eV & 0.010 & -2742 & -459.8&  0.866 & 5.111 \\ \hline
    500eV& 0.053 &-2813 & -529.3& 0.859 & 4.835  \\    \hline
    1000eV& 0.106 & -2886 & -598.5 & 0.838   & 4.148 \\    \hline
 
    \end{tabular}
 \end{table}  

\begin{figure}[h]
\includegraphics[width=0.5\textwidth]{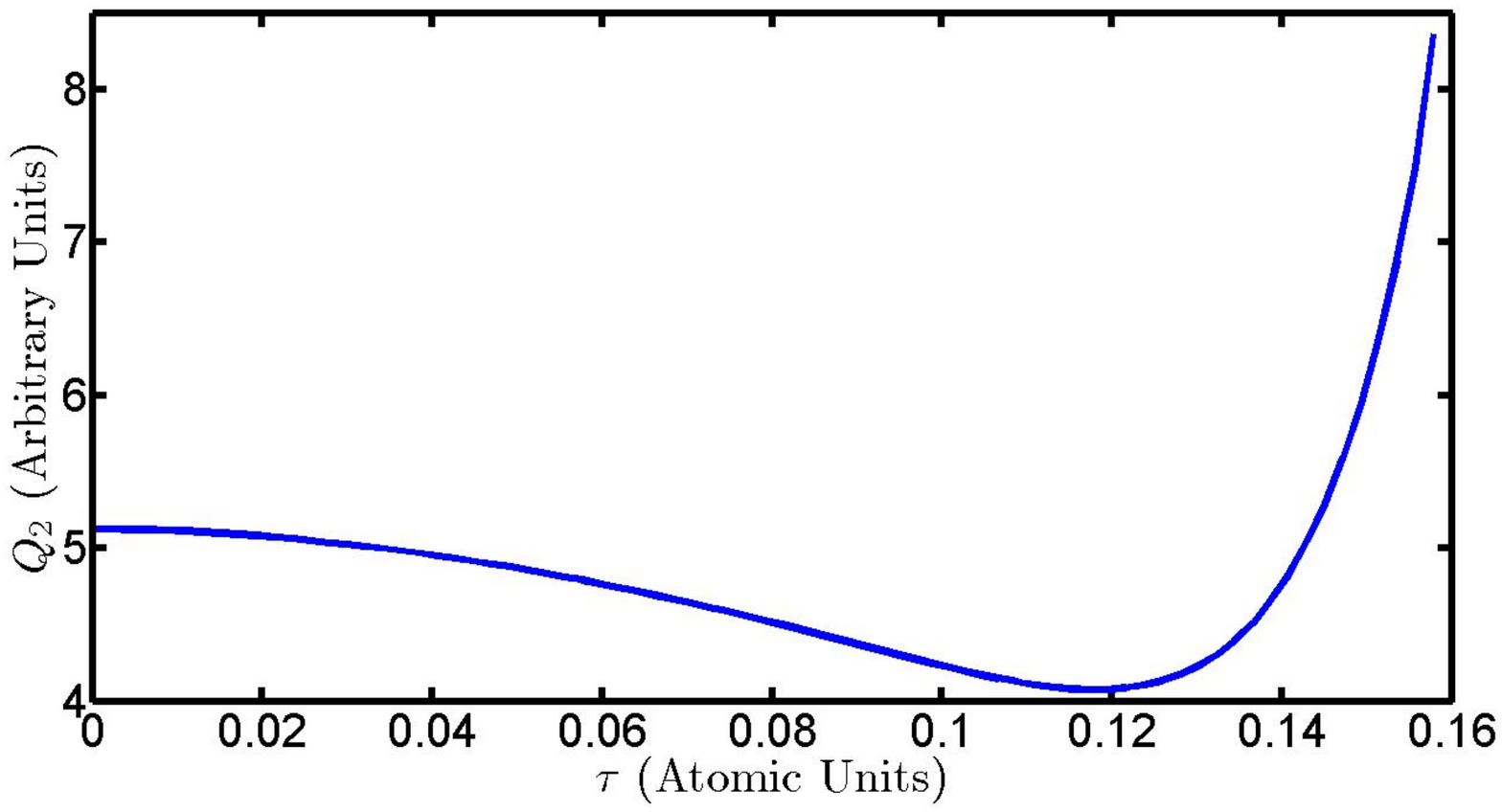}
\caption{$\tau$ dependence of $Q_{2}$. Shown are calculations for Mercury $Z=80$,
$\rho=1\frac{g}{cm^{3}}$ at $T=1-1500\ev$. \label{fig:8}}
\end{figure}

The model breaks when the condition $\mu_1<\mu_2\ll\mu$ is broken. This breaking indicates L-shell ionization, and thus a change in $Q_2$. Indeed, in Fig. \ref{fig:8} we can see that, for Mercury at $\rho=1\gcc$, $Q_{2}$ is of the same order until it reaches its minimum at $T=1250\ev$. For
higher temperatures it starts to increase rapidly, adding a growing number of quantum mechanical electrons, while not subtracting enough semi-classical ones. For even higher
temperatures $\mu_{2}$ and $Q_{2}$ do not converge. For Mercury with $\rho=1\gcc$ it occurs at about $T=1550\ev$. Thus, the behavior of $Q_2$ can be used to probe the validity of the model. Giving the strongly bound electrons a finite temperature treatment does not solve this validity problem. In appendix B we demonstrate the TFS model effect on the pressure.

\section{Conclusions}
We have  established a consistent method to estimate a $Z^{-1/3}$ correction to the TF energy at finite thermodynamic conditions, in a functional form. 
The model is valid as long as the L-shell is not ionized. We have found that the correction to the TF energy in our regime is as in the cold model, namely $\frac{1}{2}Z^{2}$.
The strongly bound electrons are not affected by the temperature in our model, because in the temperature scope we are dealing, the most inner electron are not yet affected by the environment. Nonetheless, our model give a finite temperature model that deal with the strongly bound electrons and gives a more accurate result than the zero order finite temperature TF. The model presented here takes into account the electron screening potential when dealing with the strongly bound electrons. This accounts for the differences comparing to Ref.~\cite{22,23,24,25}.

When comparing our model to KS-DFT  we see that KS-DFT deals with the strongly bound electron automatically. However, in KS-DFT deriving an order by order correction to a functional is challenging. In addition, our OFDFT model can help build better 
functionals for KS-DFT using the physical knowledge acquire in our research. Above all, our model is highly  efficient computationally with respect to KS-DFT.

The results lead us to conjecture that TFS correction is
the leading correction for the TF model at finite temperature. In particular due to the fact that gradient and exchange-correlation corrections to the TF energy are of order $Z^{\frac{5}{3}}$ \cite{29}. As such, it is essential to include the TFS correction as a starting point when studying these further corrections, as well as relativistic corrections. Such successive construction of functionals will allow to better estimate theoretical uncertainties and will be used to derive equations of state and compute opacity in WDM in a low computational effort. Furthermore, the potential of the model  can be used a pseudo-potential in Molecular Dynamics calculations. Moreover, the understanding of the importance of each correction to the relevant physical problem can help build  KS-DFT functionals that are more accurate and are relevant for the problem at hand.

In future work, we intend to investigate different modeling of the plasma environment surrounding the ions \cite{100,101,102} going beyond the ion-sphere model to corrections such as short range ion-ion and electron-ion correlations. Including such correlations becomes important as the plasma parameter grows.  In addition, we intend to extend the validity regime of the model to higher temperatures, which will demand, among other corrections, a finite temperature treatment for the strongly bound electrons. Furthermore, we will add  gradient, exchange-correlation and relativistic corrections to our model.
\section{acknowledgments}
The authors thank Samuel B. Trickey for valuable comments on earlier versions of this paper. E.\ S.\ thanks the hospitality of the University of Florida during the completion of this work.
\appendix
\section{Scaling of the TF Model with the Strongly Bound Electron Correction}
In this appendix we will attain the $Z$ scaling properties of the correction
we suggested. We use Hartree atomic unit in this part. We define
\begin{widetext}
\begin{eqnarray}
F^{(1)} & = & \frac{1}{2}\Sigma_{j=1}^{2}\int d^{3}r\frac{\sqrt{2}}{\pi^{2}\beta^{\frac{5}{2}}}\left(-\frac{2}{3}I_{\frac{3}{2}}(\beta(V+\mu))+\frac{2}{3}I_{\frac{3}{2}}(\beta(V+\mu_{j}))-\frac{1}{2}\Sigma_{j=1}^{2}\beta(\mu_{j}-\mu)I_{\frac{1}{2}}(\beta(V+\mu_{j}))\right),\label{eq:153}
\end{eqnarray}
\end{widetext}
\begin{eqnarray}
F^{(2)} & = & \int d^{3}r\left(-\frac{1}{8\pi}(\nabla(V(r)-\frac{Z}{r}))^{2}\right),\label{eq:154}
\end{eqnarray}
and

\begin{eqnarray}
F^{(3)} & = & \int d^{3}r\left(-V(r)-\mu\right)\rho_{s}(r).\label{eq:155}
\end{eqnarray}
Thus, the energy functional from Eq. (\ref{eq:140}) is

\begin{eqnarray}
F_{TFS} & = & F^{(1)}+F^{(2)}+F^{(3)}+Z^{2}n_{s}+\mu N.\label{eq:156}
\end{eqnarray}
Under the same scaling in $\lambda$ as in paper, we have the following equation

\begin{eqnarray}
\frac{d}{d\lambda}\left(F^{\lambda(1)}+F^{\lambda(2)}+F^{\lambda(3)}\right)\mid_{\lambda=1}
+2(\alpha-1)Z^{2}n_{s}&&\nonumber \\
+\mu\text{\ensuremath{\alpha}}N=(\alpha-1)Z\frac{\partial F}{\partial Z}-\alpha\beta\frac{\partial F}{\partial\beta}-r_{0}\frac{\partial F}{\partial r_{0}}.&&
\end{eqnarray}
We substitute $\alpha=4$, and use the methods from Ref. \cite{10} to estimate the derivative of $F$ with respect to $\lambda$ and get
\begin{eqnarray}
7F & = & F^{(3)}-\Sigma_{j=1}^{2}Q_{j}\left(\mu_{j}+\left\langle \frac{d}{dr}\left(rV(r)\right)\right\rangle _{n_{j}}\right)-2\mu N_{s}\nonumber \\&-&2\int d^{3}r\frac{d}{dr}\left(rV\right)\rho_{s}+Z^{2}n_{s}\nonumber \\
& + & 3\mu N+3Z\frac{\partial F}{\partial Z}-4\beta\frac{\partial F}{\partial\beta}-r_{0}\frac{\partial F}{\partial r_{0}}.\label{202}
\end{eqnarray}

We approximate the potential for our model, $V_{TFS}(r)$, as series in the spirit of Baker \cite{28} with a dependency in $Z$ besides the dependency in $\sigma$ and $\tau$ for the coefficient of $Z^{\frac{4}{3}}$:

\begin{eqnarray}
V_{TFS}(r) & = & \frac{Z}{r}-a\left(Z,\sigma,\tau\right)Z^{\frac{4}{3}}.\label{eq:159}
\end{eqnarray}
In addition, the strongly bound electrons correction does not affect the chemical potential.
Therefore, we can use the TF chemical potential from Eq. (\ref{1}).
The following approximations are a direct result of the first approximation and $\mu=cZ^{4/3}$ 
\begin{eqnarray}
\mu_{j} & = & -\frac{Z^{2}}{2n_{j}^{2}}+aZ^{\frac{4}{3}},\label{eq:160}
\end{eqnarray}
\begin{eqnarray}
\frac{\partial}{\partial r}\left(rV\right) & = & -aZ^{\frac{4}{3}},\label{eq:161}
\end{eqnarray}
\begin{eqnarray}
\int d^{3}r\frac{\partial}{\partial r}\left(rV\right)\rho_{s} & = & -aZ^{\frac{4}{3}}N_{s},\label{eq:162}
\end{eqnarray}
and
\begin{equation}
F^{(3)}=-2Z^{2}n_{s}+aZ^{\frac{4}{3}}N_{s}-cZ^{\frac{4}{3}}N_{s}.\label{eq:163}
\end{equation}

Here $ N_{s}= \int d^{3}r\rho_{s},\label{201} $ for $r_{0}\rightarrow \infty$. 
We can assume this holds for finite radius from the atom as well due to the fact that
\begin{eqnarray}
\int d^{3}r\rho_{s}  & = & \sum_{n=1}^{n_{s}}2n^{2}\left(1-4\pi\int_{r_{0}}^{\infty}drr^{2}\frac{Z^{3}}{4\pi}16exp(-4Zr)\right)\nonumber \\&=&  \sum_{n=1}^{n_{s}}2n^{2}\left(1+\frac{\left(8Z^{2}r_{0}^{2}+4Zr_{0}+1\right)}{2exp(4Zr_{0})}\right),
\end{eqnarray}
and when taking the exponent to second order we get
\begin{equation}
\int d^{3}r\rho_{s}=N_{s}=\sum_{n=1}^{n_{s}}2n^{2}\left(1+\frac{1}{2}\right).
\end{equation}

We turn now to  approximate $Q_{j}$, which can be written as
\begin{eqnarray}
Q_{j} & = & \frac{2\sqrt{2}}{\pi\beta^{\frac{1}{2}}}\left(\frac{Z^{2}}{2n_{j}^{2}}-\left(a-c\right)Z^{\frac{4}{3}}\right)h(\beta,Z),\label{eq:164}
\end{eqnarray}
where $h(\beta,Z)$ is 
\begin{eqnarray}
h(\beta,Z)  =  \int_{0}^{\frac{2n_{j}^{2}}{Z}}drr^{2}\left(I_{-\frac{1}{2}}\left(\beta\frac{Z}{r}-\frac{\beta Z^{2}}{2n_{j}^{2}}\right)\right).\label{eq:212}
\end{eqnarray}
When we use $\frac{2n_{j}^{2}}{Z}$ as the boundary of the integration
on $r$. The reason for this  estimation is that $Q_{j}$ main contribution comes from the non semi-classical area, while using $\mu_{j}$ as the semi-classical limit. In addition, we can numerically establish that for $\mu_j\ll\mu$
we get the following

\begin{eqnarray}
\frac{2\sqrt{2}}{\pi\beta^{\frac{1}{2}}}h(\beta,Z) & \sim &\frac{2n_{j}^{5}}{Z^{2}}.\label{eq:165}
\end{eqnarray}
Thus,
\begin{eqnarray}
Q_{j} & \cong & \frac{2n_{j}^{5}}{Z^{2}}(\frac{Z^{2}}{2n_{j}^{2}}-aZ^{\frac{4}{3}}+cZ^{\frac{4}{3}}).\label{eq:167}
\end{eqnarray}

Using the approximations in Eq. (\ref{eq:159},\ref{eq:160},\ref{eq:161},\ref{eq:162},\ref{eq:163},\ref{eq:167}) we get the following equation

\begin{eqnarray}
7F & = & 3cZ^{\frac{7}{3}}+\frac{1}{2}Z^{2}-\left(a-c\right)\left(2n_{s}+1\right)Z^{\frac{4}{3}}\nonumber \\&+&3Z\frac{\partial F}{\partial Z}-r_{0}\frac{\partial F}{\partial r_{0}}.
\end{eqnarray}
We solve using the characteristics method to get
\begin{eqnarray}
\frac{dZ}{3Z}=-\frac{d\beta}{4\beta}=-\frac{dr_{0}}{r_{0}}=...&&\nonumber\\...=\frac{dF}{7F+3cZ^{\frac{7}{3}}-\frac{1}{2}Z^{2}+\left(a-c\right)\left(2n_{s}+1\right)Z^{\frac{4}{3}}}.&&
\end{eqnarray}
First, we solve for $F$ and $Z$ in the spirit of Ref. \cite{10}, namely
\begin{equation}
Z\frac{dF}{dZ}=\frac{7F}{3}-cZ^{\frac{7}{3}}-\frac{1}{6}Z^{2}+\frac{1}{3}\left(a-c\right)\left(2n_{s}+1\right)Z^{\frac{4}{3}}.\label{203}
\end{equation}
We calculate $Z\frac{dF}{dZ}$
\begin{eqnarray}
Z\frac{dF}{dZ} &=&  \frac{1}{2}\sum_{j=1}^{2}Q_{j}\left(\mu_{j}+\frac{Z^{2}}{2n_{j}^{2}}+e\left\langle \frac{d}{dr}\left(rV(r)\right)\right\rangle _{n_{j}}\right)\nonumber\\
&+&Z\left(V-\frac{Z}{r}\right)\mid_{r=0}+F^{(3)}+\mu N_{s}\nonumber\\&+&\int d^{3}r\frac{d}{dr}\left(rV\right)\rho_{s}.
\end{eqnarray}
Using the same approximation we  get
\begin{equation}
Z\frac{dF}{dZ}=-aZ^{\frac{7}{3}}.
\end{equation}
Putting this in Eq. (\ref{203}) gives
\begin{equation}
7F=-3(a-c)Z^{\frac{7}{3}}+\frac{1}{2}Z^{2}-\left(a-c\right)\left(2n_{s}+1\right)Z^{\frac{4}{3}}.\label{204}
\end{equation}
We take the derivative of Eq. (\ref{204}) and multiple it by $Z$ to get
\begin{widetext}
\begin{equation}
-(a-c)Z^{\frac{7}{3}}=Z\frac{d}{dZ}\left(-\frac{3}{7}(a-c)Z^{\frac{7}{3}}+\frac{1}{14}Z^{2}-\frac{1}{7}\left(a-c\right)\left(2n_{s}+1\right)Z^{\frac{4}{3}}\right).
\end{equation}
\end{widetext}
We solve this equation for $\left(a-c\right)$ 
\begin{eqnarray}
a-c & = & A(\sigma,\tau)\left(1+\frac{2n_{s}+1}{3Z}\right)^{\frac{4}{3}}\nonumber\\ &-&\frac{1}{Z^{\frac{1}{3}}}\left(1+\frac{2n_{s}+1}{4Z}\right),
\end{eqnarray}
when $A$ is a universal coefficient.
Putting this back in Eq.(\ref{204}) we get
\begin{eqnarray}
F &=&  -\frac{3}{7}A(\sigma,\tau)\left(1+\frac{2n_{s}+1}{3Z}\right)^{\frac{4}{3}}Z^{\frac{7}{3}} \nonumber\\&+&\frac{1}{2}Z^{2}\left(1+\frac{6n_{s}+3}{28Z}\right)+O\left(Z^{\frac{4}{3}}\right)
\nonumber\\&=&  -\frac{3}{7}A(\sigma,\tau)Z{}^{\frac{7}{3}}+\frac{1}{2}Z^{2}+O\left(Z^{\frac{4}{3}}\right).\label{eq:175}
\end{eqnarray}

Finally we have 
\begin{equation}
\frac{F}{-\frac{7}{3}A\left(\sigma,\tau\right)Z^{\frac{7}{3}}+\frac{1}{2}Z^{2}+O(Z^{\frac{4}{3}})}=const.
\end{equation}
Together with the results from the TF model we deduce
\begin{equation}
F_{TFS}=g_{s}(\sigma,\tau)\left(-\frac{7}{3}A\left(\sigma,\tau\right)Z^{\frac{7}{3}}+\frac{1}{2}Z^{2}+O(Z^{\frac{4}{3}})\right),
\end{equation}
which can be can understood as
\begin{equation}
F_{TFS}=g(\sigma,\tau)Z^{\frac{7}{3}}+g_{s}(\sigma,\tau)\frac{1}{2}Z^{2}+O(Z^{\frac{4}{3}}).\label{eq:208}
\end{equation}
By the numerical results we see that $g(\sigma,\tau)=f(\sigma,\tau)$, which is the universal function for the TF model in finite temperature that is defined in Eq. (\ref{eq:20}), and that TFS universal function $g_{s}(\sigma,\tau)$ is constant with $g_{s}(\sigma,\tau)=1$. This leads us to
\begin{equation}
F_{TFS}=f(\sigma,\tau)Z^{\frac{7}{3}}+\frac{1}{2}Z^{2}+O(Z^{\frac{4}{3}}),\label{eq:209}
\end{equation}
which is Eq. (\ref{eq:122}). 

In order to fit Fig. \ref{fig:10} we compute the coefficients of the lower term in the $Z^{1/3} $ expansion of Eq. (\ref{eq:175}). We do so by using $n_s=1$ and the expansion of $\left(1+\frac{2n_{s}+1}{3Z}\right)^{\frac{4}{3}}$ around infinity. namely,
\begin{eqnarray}
F_{TFS}&=&f(\sigma,\tau)Z^{\frac{7}{3}}+\frac{1}{2}Z^{2}+\frac{3}{7}f(\sigma,\tau)Z^{\frac{4}{3}}\nonumber\\&-&\frac{3}{4}Z+O(Z^{\frac{1}{3}}).\label{eq:210}
\end{eqnarray}

\section{Equation of State}
\begin{figure}[h]
	\includegraphics[width=0.5\textwidth]{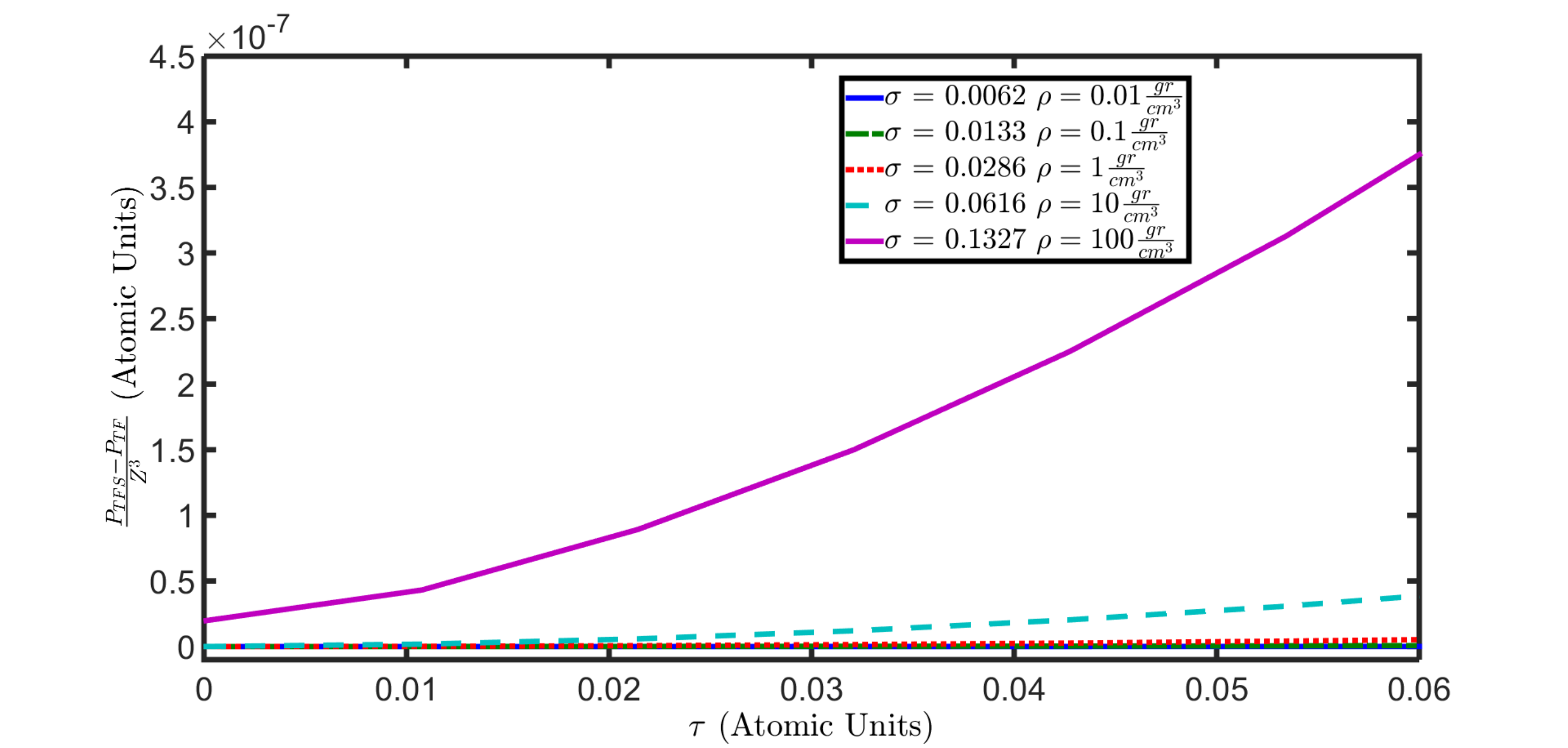}
	\caption{Dependence of the  TFS correction to pressure on $\tau$ scaled in $Z$. Shown are calculations for Mercury $Z=80$
		at $\rho=0.01,0.1,1,10,100\frac{gr}{cm^3}$, and $\sigma=0.0062,0.0133,0.0286,0.0616,0.1327$. \label{fig:11}}
\end{figure}

As an example for the effect the TFS correction has on the observables, we present in Fig. \ref{fig:11} the predicted correction to the TF pressure due to the strongly bound electron correction. The analytic form for the suggest model is
\begin{widetext}
\begin{eqnarray} 
P_{TFS} & = & \frac{2\sqrt{2}}{3\pi^{2}\hbar{}^{3}}\frac{m^{\frac{3}{2}}}{\beta^{\frac{5}{2}}}\left(I_{\frac{3}{2}}(\beta\mu)\right) +\left(\mu-\frac{e^{2}Z}{r_{0}}\right)\rho_{s}(r_{0}) -\Sigma_{j=1}^{2}\frac{m^{\frac{3}{2}}\sqrt{2}}{2\pi^{2}\hbar^{3}\beta^{\frac{5}{2}}}\left(\frac{2}{3}I_{\frac{3}{2}}(\beta\mu_{j})-\beta(\mu_{j}-\mu)I_{\frac{1}{2}}(\beta\mu_{j})\right).
\end{eqnarray}
\end{widetext}
Similarly to the energy in Eq. (\ref{eq:209}), it can be analytically and numerically shown that the TFS correction to the pressure is only $Z^{-1/3}$ suppressed with respect to the leading TF contribution.

\end{document}